# LITHIUM ANTINEUTRINO SOURCE IN THE TANDEM SCHEME OF THE ACCELERATOR AND NEUTRON PRODUCTING TUNGSTEN TARGET

V. I. Lyashuk [1,2]

[1] *Institute for Nuclear Research, Russian Academy of Science, Moscow, Russia*
[2] *National Research Center "Kurchatov Institute", Moscow, Russia*

Abstract:
The antineutrinos of the neutron rich $^8$Li isotope is characterized by hard and good defined spectrum - averaged energy $E_{\tilde{\nu}}$ = 6.5 MeV and maximal - up to 13 MeV. An intensive antineutrino source with such parameters can be unique instrument for neutrino investigations and especially for search of sterile neutrinos. The $^8$Li can be produced by (n,γ)-activation of $^7$Li isotope. The proposed scheme of the antineutrino source is based on the lithium blanket around the accelerator neutron producting target. We propose to use heavy water solution of the lithium hydroxide instead of lithium in metallic state. Such solution for lithium blanket substance ensure the large perspectives in real steps for creation of this installation.
An analyses of neutron fields in the blanket and distribution of $^8$Li creation allows to propose the next principal steps in the construction of the lithium blanket. We propose to enclose the blanket volume isolating it's central part with more high $^8$Li production. This solution allows to decrease the necessary mass of very pure $^7$Li isotope (with admixture of $^6$Li - 0.0001) up to 120-130 kg (i.e., in ~ 150 times compare to variant of lithium blanket in metallic state). The source become more compact - the linear dimension is decreased in 2.5 times (up to 1.3 m) and this is very essential for sterile neutrino search taking in mind the indication on $\triangle m^2$~1eV$^2$ refer to active neutrinos.

## 1. Introduction

Up today the nuclear reactor is the most used electron antineutrino source. But the known advantages (high flux and avalibility) are in serious conflict with it's disadvantages. They known too. The reactor $\tilde{\nu}_e$-spectrum is sharply dropping (this fact is very negative for detection of threshold reactions - registration of the reactions with deuteron in neutral and charged channels). The energy of the reactor spectrum is lower than 10 MeV. The resulting reactor antineutrino spectrum (which is formed by β$^-$-decay of fission fragments of four main fuel isotopes - $^{235}$U, $^{239}$Pu, $^{238}$U, $^{241}$Pu) varies significantly during the reactor operation period (about 300 days). The errors of the partials spectrum (of fuel isotopes) rise very significantly with increase of the antineutrino energy. The errors rise up to: 30% at 7.5 MeV for $^{238}$U, 56% at 9.5 MeV for $^{235}$U; 80% at 8.5 MeV for $^{239}$Pu; 90% at 9.0 MeV for $^{241}$Pu [1 - 3].

In contrast to reactor spectrum the energy of $^8$Li $\tilde{\nu}_e$-spectrum is larger - $E_{\tilde{\nu}}^{\max}$ = 13 MeV and well defined. The main part of the spectrum is distributed above the thresholds of the channels for ($\tilde{\nu}_e$,d)-reaction ($E_{threshold}$ = 2.53 МэВ in neutral channel: $\tilde{\nu}_e + d \rightarrow n + p + \tilde{\nu}_e$; $E_{threshold}$ = 4.0 MeV for charged channel: $\tilde{\nu}_e + d \rightarrow n + n + e^+$). The last fact is of prime importance as at the considered MeV-energies the cross section of the neutrino interaction is proportional to squared energy: $\sigma \sim E_\nu^2$. The idea to use $^8$Li decay as a source of antineutrinos was suggested by L. A. Mikaelian, P. E. Spivak. and V .G. Tsinoev [4].

## 2. Physical principles for creation of the lithium antineutrino source

The most simple solution for construction of the $^8$Li antineutrino source is to arrange the lithium blanket close to the intensive neutron source and to ensure an effective activation - $^7$Li(n,γ)$^8$Li. In this case the lithium blanket "acts" as converter of neutrons to antineutrinos. Active zone of the reactor is respective intensive neutron source. The questions of $\tilde{\nu}_e$-source based on the reactor active zone were discussed in the works [5 - 7].

The serious requirements arise to the purification of the lithium: natural lithium consists of two isotopes: $^7$Li (92.5%) and $^6$Li (7.5%). Isotope of $^6$Li is a strong neutron absorber -$\sigma_a \approx$ (937-940) b in the thermal point. On the contrary the cross section $\sigma_{7Li(n,\gamma)} \sim$ 45 mb. So, in order to ensure the high efficiency $k$ of the blanket (number of $^8$Li isotope creation per neutron escaped from the source) the natural requirement is to decrease concentration of $^6$Li up to ~ 0.0001 (and not more than 0.0002). The calculations shows that the necessary mass of $^7$Li (of 0.9999 purity) must be ~ 20 tons and more [5-7]. The principal solution of the large $^7$Li mass problem is to use deuterium chemical compounds of $^7$Li. Let us do the next important step - we will use heavy water solution of deuterium compounds of $^7$Li (namely $D_2O$-solution of lithium hydroxide). The questions of lithium blankets on the base of deuterium compounds of $^7$Li and their heavy water solutions are discussed in [8, 9]. Namely heavy water solution of lithium hydroxides (LiOD, LiOD·$D_2O$) allows to decrease the requested mass of $^7$Li in tens times and more.

Since the 70-th years of the XX-th century the possibility to use the spallation reaction for creation of the powerful neutron source began attract more and more attention and were realized in a number of scientific centers of the Europe, Russia, Japan, USA and are developing: IREN, IFMIF, JSNS/J-Park, ESS, CSNS; project of electronuclear installation "Energy amplifier" by C. Rubbia, et al. The large benefits of spallation neutron source (usually made from tungsten, lead, tantalum, uranium, mercury and beryllium (as neutron reflector and breeder) is caused by the fact that as the proton energy is increasing the neutron yield $Y_n$ (per proton) is increasing sharply: so, for heavy targets and proton energy $E_p$ = 200-300 MeV the neutron yield is about $Y_n \approx$ (1.6-3.5), for $E_p$ =500-600 MeV the yield increases up to $Y_n \approx$ 10.

### 3. Effective scheme of the lithium antineutrino source on the base of accelerator

In accelerator variants for creation of intensive neutrino source we proposed to surround the target by lithium converter [10]. It was investigated the accelerator variant of the lithium antineutrino source where the lithium blanket surrounded the tungsten target (bombarded by proton of energy 50-300 MeV) [11-12]. The discussed blanket has the cylindrical geometry with 3.4 m in length, radial thickness – 1.7 m and central target (cooled by heavy water). For the blanket substance we also used heavy water solution of $^7$LiOD with concentration 9.5%. This lithium blanket was proposed as $\tilde{\nu}_e$-source for search of sterile neutrino with use of the constructed now JUNO neutrino detector [13]. The obtained results reveals that sensitivity to the mixing-angle may be extremely high: $\sin^2(2\theta) \leq 0.001$ for $\Delta m^2_{41} \gtrsim 0.2$ eV$^2$ in the model (3+1) - three active plus one sterile neutrino [14]. This simulation of the possible sterile neutrino oscillation requested to take into account the distribution of $^8$Li yield in the blanket [14]. For this purpose the blanket was divided into 10 equal steps in length and 10 equal layers in radius. The summary number of created cells was 105 ones including five nonstandard cells after the target. The scheme of the cell geometry in the blanket is given in the Fig.1.

In the calculation [14] we assume the proton energy $E_p$=200 MeV, $^6$Li admixture - 0.0001, and purification of deuterium for heavy water solution was taken into account as : D – 0.99 and $^1$H – 0.01. The obtained $^8$Li yield in the cells (see Fig. 2) confirmed the think to diminish the size of the neutrino source by separating of the central volume (with the more high yield) from outer volume (free of $^7$Li isotope). In the Fig. 2 you can ascertain that main $^8$Li yield is concentrated in the central volume around the target. The histograms values are normalized per unit. In the Fig. 1 we marked out three volumes: 1) with $^8$Li yield - 68%; 2) with $^8$Li yield - 80%; 3) with $^8$Li yield - 90%. The cell histograms for corresponding cell volumes are presented in the Fig. 2.

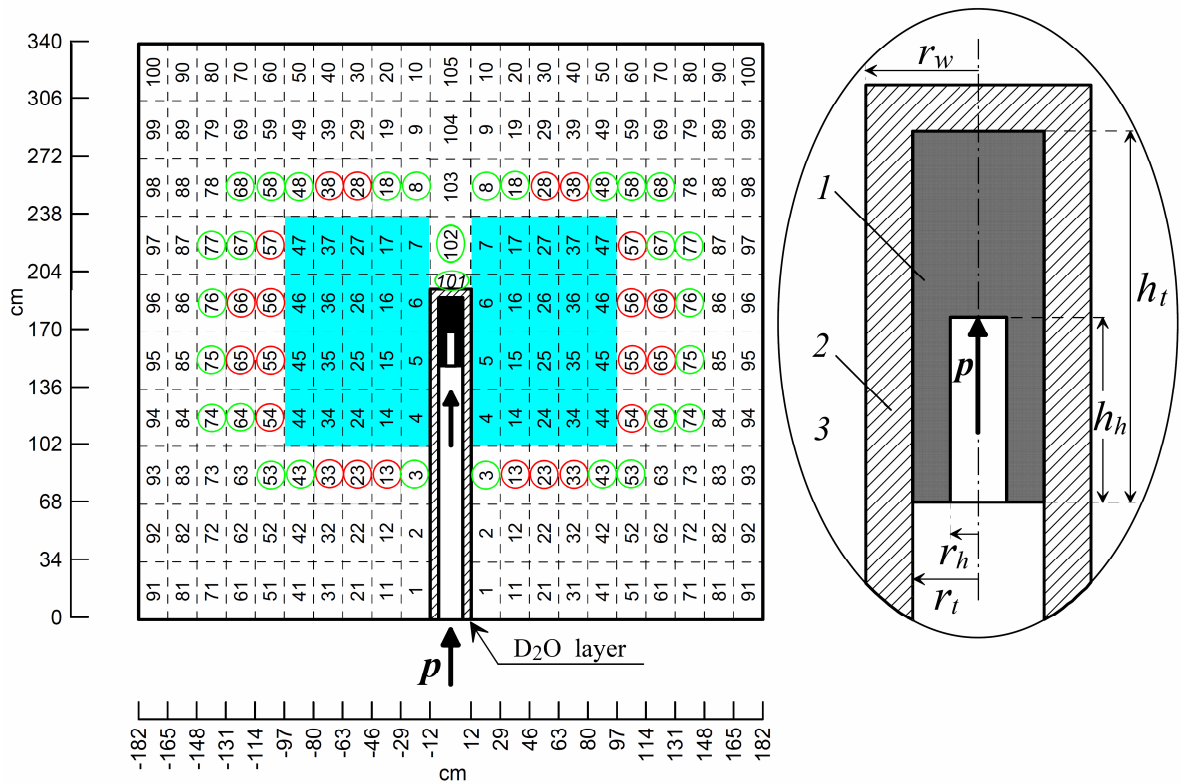

Fig. 1. The scheme of the lithium blanket with central target. The volume of the blanket is divided in 105 numbered cells. The blue cells correspond the 8Li yield - 68%. The cells in the red circles plus blue cells correspond the $^8$Li yield - 80%. The cells in the green circles plus cells in the red circles plus blue cells correspond the 8Li yield - 90%. The scheme of the target is given on the right: 1 - tungsten target; 2 - D$_2$O channel for cool of the target; 3 - lithium blanket. The target dimensions see in [11, 12].

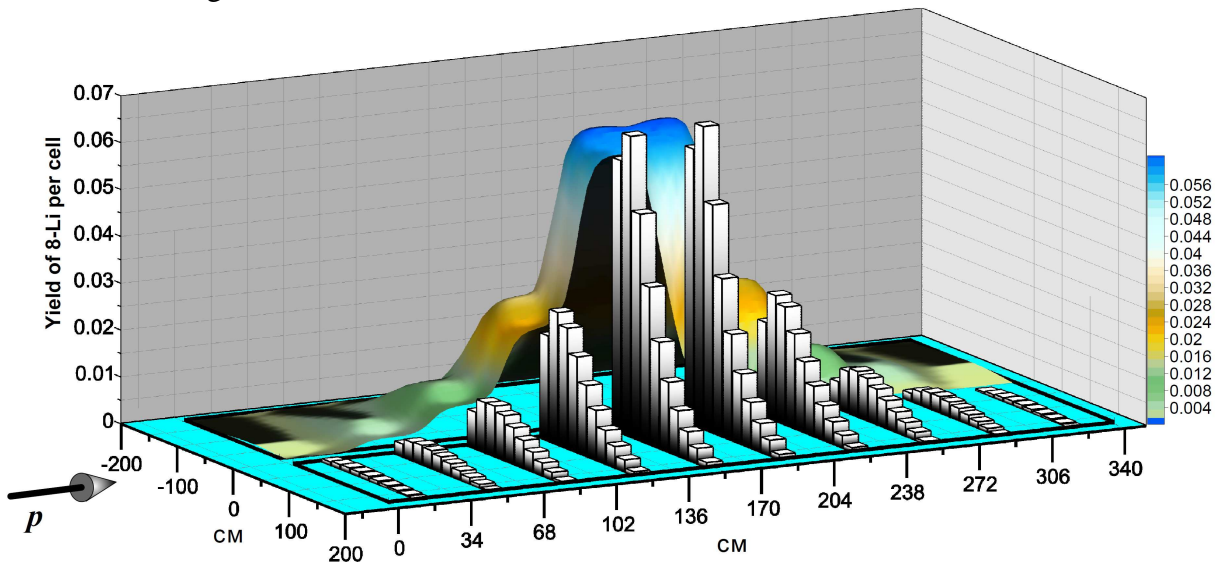

Fig. 2. The yields of $^8$Li in the cells (the histograms are on the right to the proton beam) for proton energy 200 MeV. Left to the proton beam - $^8$Li yield smoothed by spline (see colour bar on the right). The yields of $^8$Li are normalized per proton.

The $^8$Li yield has maximum in the cels 15 and 16 (Fig. 1 and Fig. 2). If to compare the density of $^8$Li distribution in the volume we can found that the maximal density of $^8$Li yield - in the cells 5 and 6 (see Fig. 3). The result is clear: this is "play" of the volume factor (geometrical factor) multiplied by density yield. The $^8$Li density yield is given in the Fig. 3 (in linear scale presentation) and also in the Fig.4 in Log-scale presentation.

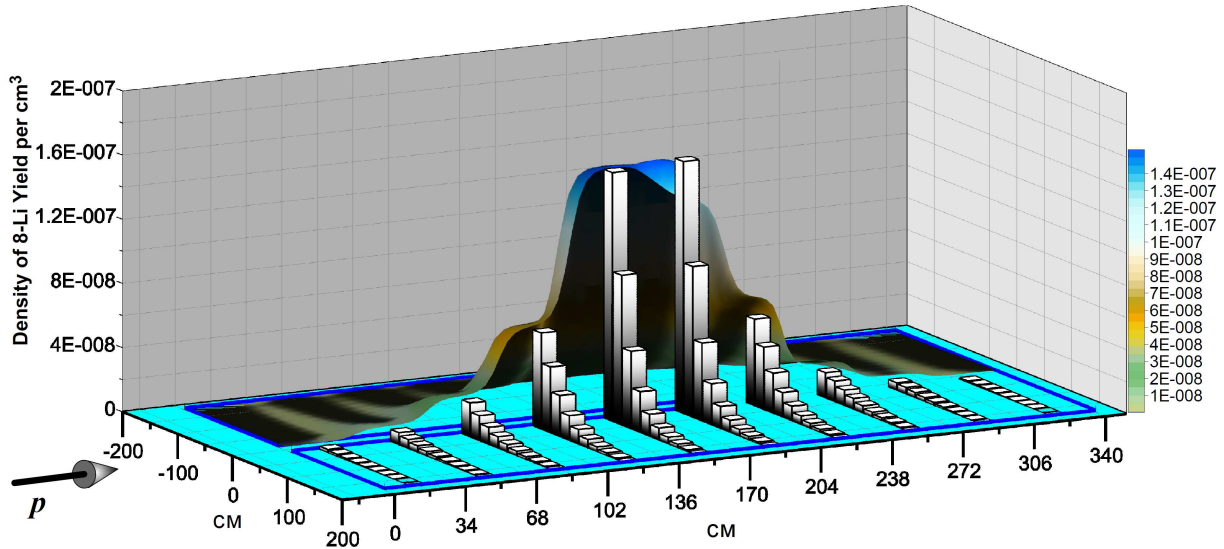

Fig. 3. The density of $^8$Li creation [1/cm$^3$] in the cells in linear scale (the histograms are on the right to the proton beam) for proton energy 200 MeV. Left to the proton beam - $^8$Li density creation smoothed by spline (see colour bar on the right). The densities of $^8$Li creation are normalized per proton.

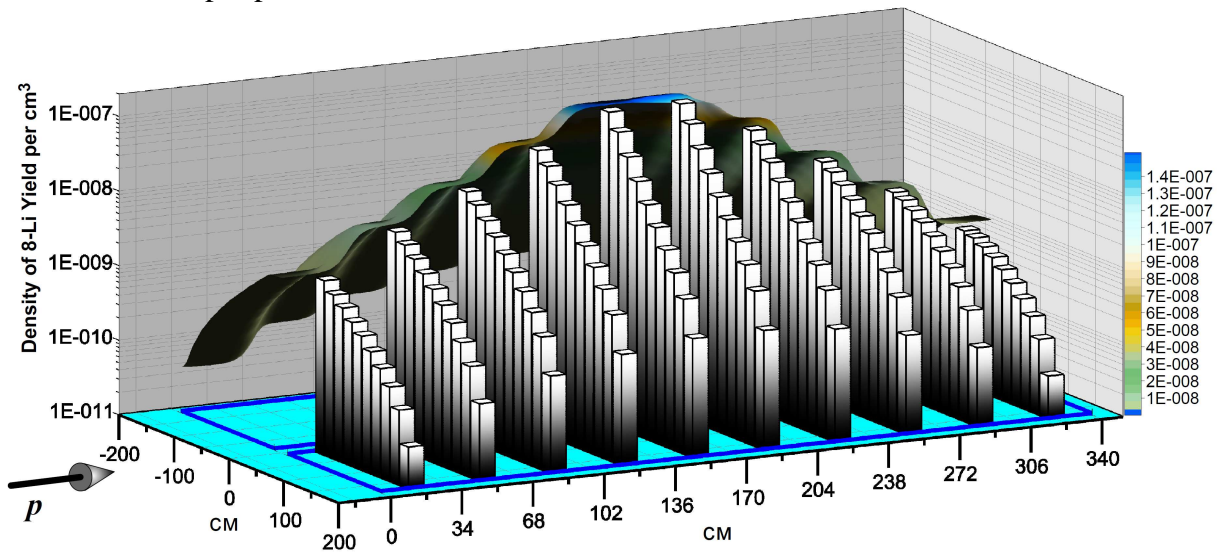

Fig. 4. The density of $^8$Li creation [1/cm$^3$] in the cells in logarithmic scale (the histograms are on the right to the proton beam) for proton energy 200 MeV. Left to the proton beam - $^8$Li density creation smoothed by spline (see colour bar on the right). The densities of $^8$Li creation are normalized per proton.

The Fig. 5 combined two graph ($^8$Li yield in the cells and density of $^8$Li creation) in the contour map. The left part - is the level lines for the smoothed $^8$Li yield in the cells, on the right - density of $^8$Li yield. The volume with large yield in the cells occupies more significant volume in the left part compare to the density distribution on the right.

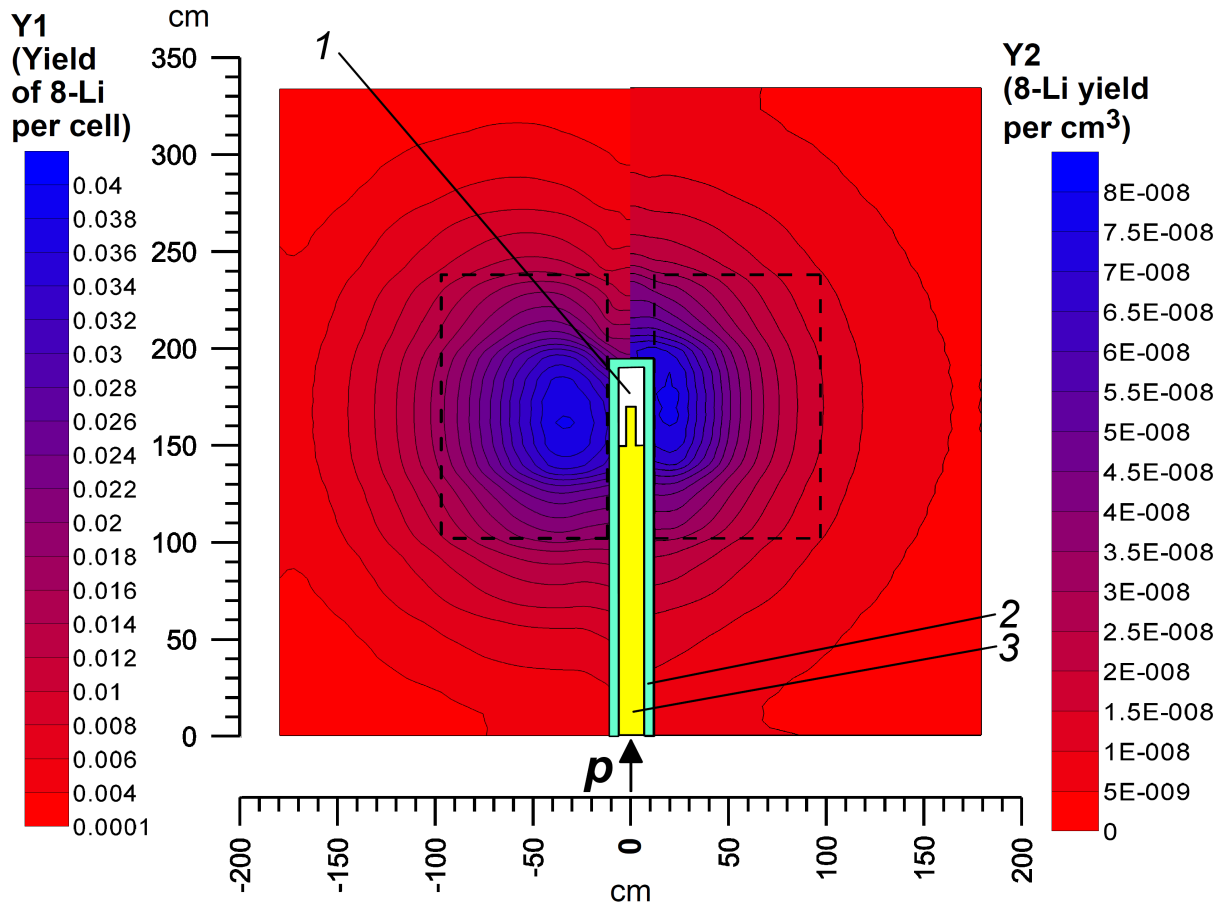

Fig. 5. Yield of $^8$Li in the cylindrical blanket (horizontal axis - the size in radius; vertical dimension - the cylinder axis). Left to the proton beam - $^8$Li normalized yield in cells - see Y1 axis. On the right - density of $^8$Li creation - see Y2 axis. 1 - target. 2 - D$_2$O channel for target cooling. 3 - channel of the proton beam. The dotted line shows the region corresponding 68% of $^8$Li yield. The yields and densities of $^8$Li creation are normalized per proton.

For realization of the task to diminish the lithium source we enclose the volume with 68% $^8$Li yield by carbon layer with thickness $L_{carbon}$. The outer volume (with respect to carbon layer) will be filled with water which will act as reflector and moderator similar to D$_2$O solution of LiOD. In the simulation we investigated the $^8$Li yield as function of the $L_{carbon}$. Let us consider the variants: 1) the $^8$Li yield for two grades of deuterium purity - 0.999 (i.e., $^1$H is

– 0.001) and 0.990 ($^1$H – 0.010); 2) possibility to change the D$_2$O to H$_2$O in the outer blanket part (see Fig. 6).

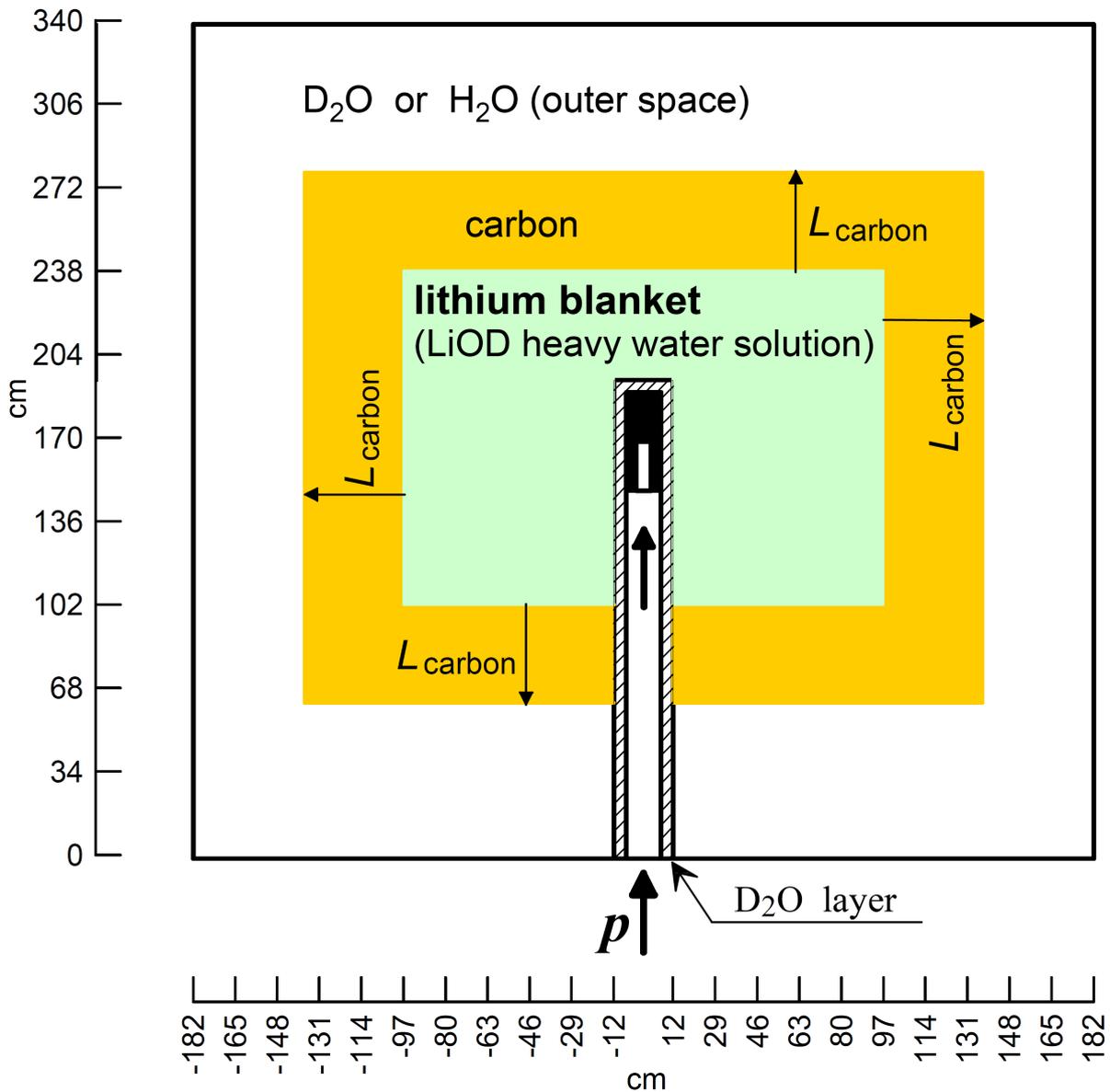

Fig.6. Scheme of the lithium blanket with carbon layer around the lithium blanket. $L_{carbon}$. - carbon layer of variable thickness. The outer space is fiiled by D$_2$O or H$_2$O.

For case of the outer D$_2$O volume the $^8$Li yield demonstrates weak dependence from the $L_{carbon}$ thickness (for both cases of deuterium purity) - see Fig. 7. But if to change outer D$_2$O part by H$_2$O the absorption on the hydrogen $^1$H will be significant for thin carbon layers $L_{carbon}$. But for larger thickness $L_{carbon} \gtrsim 50$ cm the $^8$Li yields for outer D$_2$O and H$_2$O parts are practically equal as for 0.990 as 0.999 grades of deuterium purity. Such development of the blanket geometry gives decrease of $^7$Li mass in 8.9 times compare to initial blanket volume (up to 120-130 kg) and shortens the blanket length in 2.5 times (up to 1.36 м).

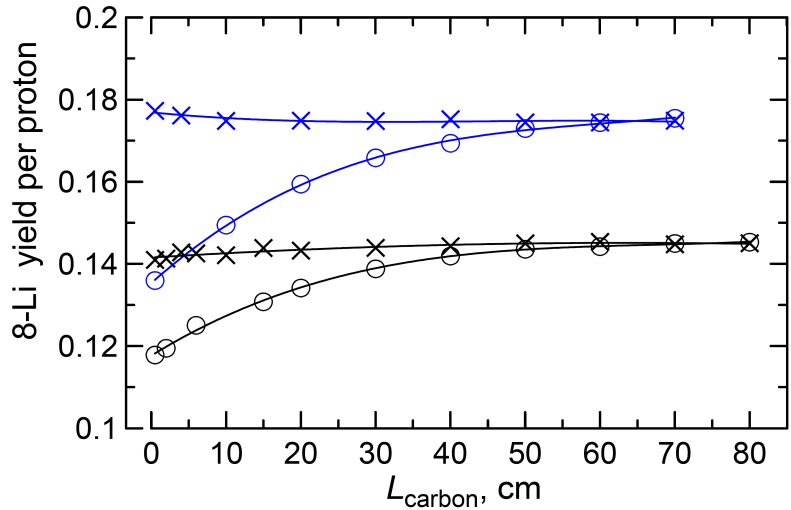

Fig.7. Dependence of $^8$Li yield on carbon layer $L_{carbon}$. Blue fit curves correspond to D-purity 0.999, black curves - D-purity - 0.990. × - the outer space is filled with $D_2O$. Circles - the outer space is filled with $H_2O$. The results are normalized per proton.

The another advantage of the such modification is the strong decrease of the neutron flux escaped from boundary of the installation - see Fig. 8. So, for $L_{carbon}$. ~ 50 cm the number of neutron escaped from the installation for $H_2O$ variant is smaller in 40-60 times compare to $D_2O$ case. In additional we can save significant $D_2O$ volume and use the light water.

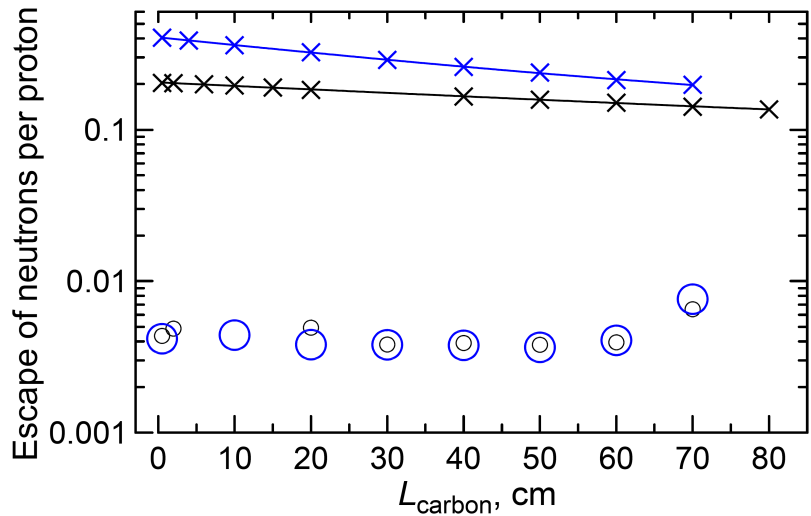

Fig.8. Dependence of neutron escaped (from boundary of the installation) on carbon layer $L_{carbon}$. Blue fit curves correspond to D-purity 0.999, black curves - D-purity - 0.990. × - the outer space is filled with $D_2O$. Circles - the outer space is filled with $H_2O$. The results are normalized per proton.

## 4. Conclusion

Let us underline once more the most important results of the discussed development of the blanket. The lithium blanket became more compact. If to use lithium in metallic state we will need ~19500 kg of pure $^7$Li. But: 1) exchange of metallic lithium to heavy water 9.5% solution of LiOD and 2) (the discussed here) diminishing of the blanket ensure decrease of the 7Li mass up to (120-130) kg.

Some fitting results of the registrated anomaly in the neutrino oscillation experiments indicate on the possibility that $\Delta m^2$ for proposed sterile neutrinos can be in the region ~ 1 eV$^2$ [15]. The results of the work [14] indicate that the possible oscillation length in the scheme (3+1) is about 7.3 m and in the scheme (3+2) - ~10.6 m. So, the obtained downsizing of the lithium source is very important for the precision of the proposed experiments [14] on sterile neutrino search.

**Acknowledgments**

The author tender thanks L. B. Bezrukov, B. K. Lubsandorzhiev Yu. S. Lutostansky and I. I. Tkachev for for helpful discussion and interest to this work.